\documentclass[a4paper]{jpconf}
\usepackage{graphicx}
\begin{document}
\title{New Heat-Capacity Measurements of the Possible Order-Disorder Transition in the 4/7-phase of 2D Helium}

\author{S Nakamura, K Matsui, T Matsui and Hiroshi Fukuyama}

\address{Department of Physics, Graduate School of Science, The University of Tokyo, 7-3-1 Hongo, Bunkyo-ku, Tokyo 113-0033,  Japan}

\ead{nakamura@kelvin.phys.s.u-tokyo.ac.jp, hiroshi@phys.s.u-tokyo.ac.jp}

\begin{abstract}
We have developed a new heat-capacity measuring system with ZYX graphite that is known to have much better crystallinity than Grafoil and started data collection.
We report preliminary data as well as a detailed description of instrumentation
including a mechanical heat-switch operated by hydraulic pressure of superfluid $^{4}$He.
\end{abstract}

\section{Introduction}
Helium atoms in the second layer on graphite are supposed to localize at the 4/7 commensurate density with respect to the first layer. 
Its low temperature magnetic behavior is well studied with various experimental techniques in the case of $^{3}$He 4/7 phases. 
These experiments claim that $^{3}$He 4/7 phase is a highly frustrated quantum spin system and its ground state is considered to be a gapless spin liquid~\cite{hf}.
Towards a further understanding of its curious property, knowledge of structural properties is required. 
However, little is known about the expected order-disorder transition at high temperatures.

Previous heat-capacity measurements revealed anomalies with a maximum at $T \sim 1.0$ K and 
$1.5$ K for $^{3}$He and $^{4}$He, respectively~\cite{grey}\cite{sv}. 
If these anomalies are originated from the same phenomena, heat capacities should be expressed as the same function of ${T}/{T_{\mathrm{peak}}}$.
Though the anomaly of $^{3}$He is 10\% smaller than expected from that of $^{4}$He, general shapes resemble each other.
The anomalies were not sharp enough, which presumably were broadened by the poor 
crystallinity of exfoliated graphite substrate (Grafoil) used in those measurements,
as in the case of the $\sqrt{3}\times\sqrt{3}$ phase in first layer~\cite{root3} 
and that makes it hard to distinguish whether those anomalies represent order-disorder transition or 
some specific energy-scale.
Moreover, in the case of $^{4}$He, the recent first-principles calculation claimes the absence of
the 4/7 phase~\cite{corboz}.

To unmask this controversy, we have developed a new heat-capacity measuring system 
with ZYX graphite which is known to have much better crystallinity than Grafoil.
ZYX graphite has magnitude larger platelet size, about $100-200$~nm, and five-times smaller mosaic angle spread ($\sim \pm 5^{\circ}$)~\cite{niimiRSI}\cite{takayoshi} than Grafoil.
Previous experiments of 2D helium on ZYX graphite are limited, especially of the 2nd layer~\cite{jinshan}. 
However, 2D helium on ZYX graphite is much promising system to study the phase transitions in detail
and is expected to be similar to that on Grafoil in phase diagram. 

\section{Experimental Setup}
\begin{figure}[htbp]
\begin{center}
\includegraphics[width=0.9\textwidth]{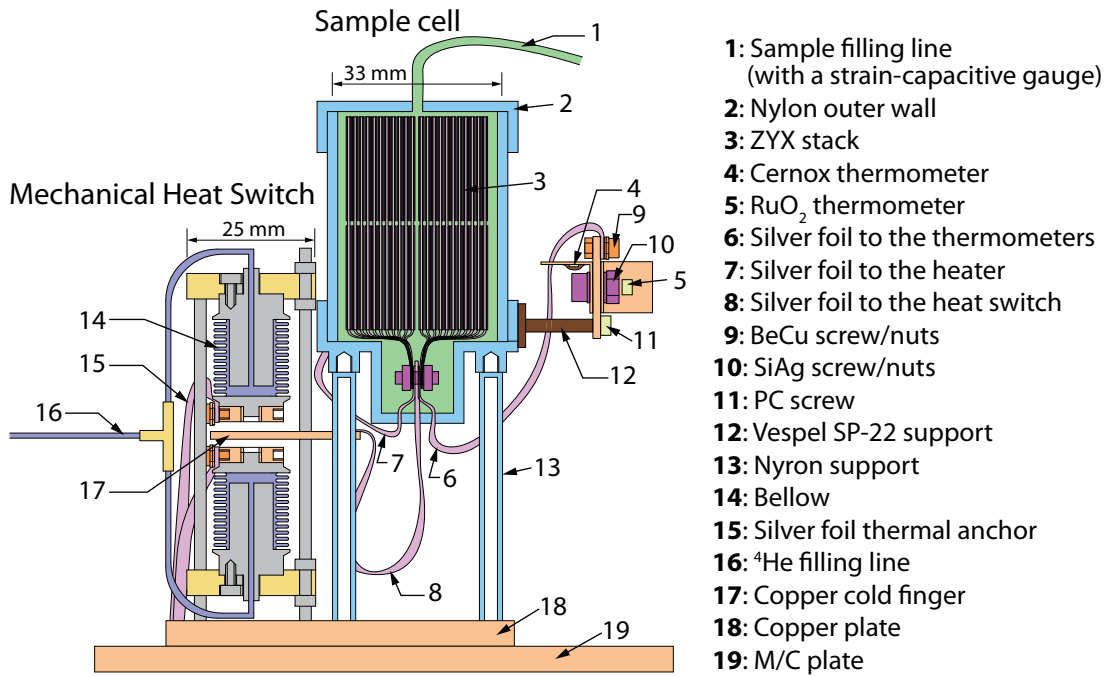}
\caption{\label{fig:cell}Calorimeter with ZYX substrate.}
\end{center}
\end{figure}

%
The experimental setup for our heat capacity measurements is shown in Figure \ref{fig:cell}.
We use a cryogen-free dilution refrigerator (Oxford Instruments, DR-200) to cool down our calorimeter.
An interior of a nylon cell is filled with a stack of ZYX exfoliated graphite.
ZYX exfoliated graphite is fabricated from a piece of ZYH grade HOPG.
We compressed ZYX graphite pieces to $1.1$~g/cc to maximize the specific surface area and cut them into 1-mm slices. 
The resultant ZYX slices were bonded onto both sides of 50~$\mu$m thick silver foils in vacuum with appropriate heat and pressure,
 then we stacked them neatly.

The temperature of 2D helium is measured by means of a resistance thermometers, 
Cernox\texttrademark (Lakeshore, CX-1030-BR) and RuO$_{2}$ (ALPS, 470~$\Omega$), 
and regulated with an PtRh$_{13\%}$ wire heater (53~$\Omega$) connected to the ZYX stack with annealed 0.1 mm-thick silver foils (RRR~$\sim 930$).
Both thermometers are calibrated against other calibrated thermometers, 
a Cernox\texttrademark (Lakeshore, CX-1050) calibrated in $1.4$~-~$300$~K, a Rox\texttrademark (Lakeshore, RX-202A)  in $0.05$~-~$40$~K and a $^{60}$Co nuclear orientation thermometer in $10$~-~$100$~mK.

The sample pressure is monitored with a strain-capacitive pressure gauge located in the vicinity of the cell. 
The diaphragm was made of hard-silver and the pressure resolution was smaller than 0.5~Pa at $T\sim 70$~K.

The adiabatic heat-pulse method requires a mechanism to switch between thermally conductive and non-conductive modes.
It is generally difficult to carry out heat capacity measurements in a wide temperature range between a few tens millikelvin and 4~K 
with only one heat switch.
Superconducting heat switch shows high switching ratio only below $0.1T_{c}$, 
gas heat switch can show enough thermal conductivity in conductive mode only above $0.1$~K even with $^{3}$He gas 
and mechanical heat switch generates too large heat ($\sim 2~\mu$J~\cite{greyMHSW}) for experiments at mK temperatures 
when it switches to non-conductive mode.

We constructed a newly designed mechanical heat switch which generate only 0.7~$\mu$J ~\cite{tsuji}
and the lower temperature limit for our measurements is estimated as $50$~mK with it.
A mechanism of the heat switch is as follows.
Two copper plates at the ends of two identical bellows wedge a copper cold finger, 
which is thermally connected to the ZYX substrate, from both sides to close the switch. 
All of the copper pieces are plated with gold.
The bellows are actuated hydraulically by liquid $^{4}$He in a pressure range between 0.1 and 2.0~MPa. 
The switch opens in reducing and closes in increasing the $^{4}$He pressure at $\sim 0.4$~MPa.
The forces produced by the bellows are canceled out and we can apply large pressing force ($<300$~N) 
to obtain high thermal conductivity in the closed mode. 
The measured electrical contact resistance in the closed mode was $73~\mu\Omega$ at 4.2~K with $^{4}$He pressure $0.5$~MPa 
and the thermal conductance is estimated as $1.4$~mW/K using the Wiedemann-Franz law.

\begin{figure}[htbp]
\begin{minipage}{0.7\textwidth}
\includegraphics[width=0.95\textwidth]{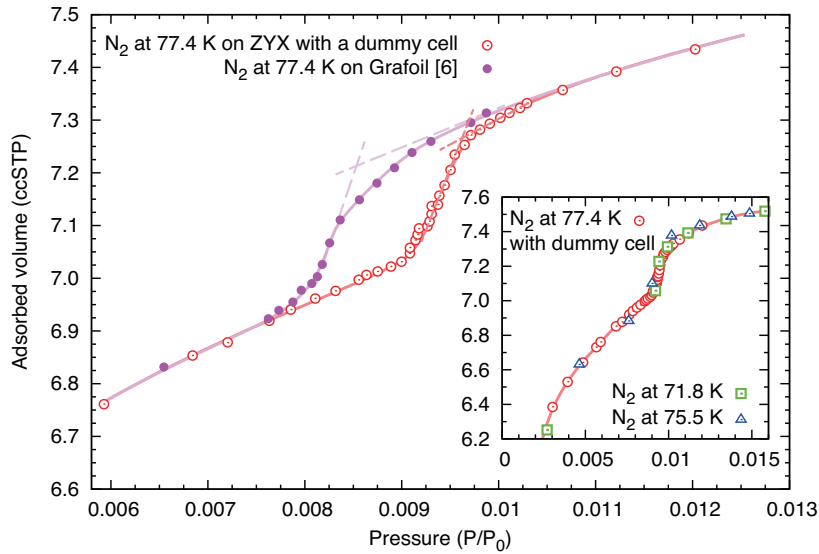}
\end{minipage} 
\begin{minipage}{0.3\textwidth}
\caption{\label{fig:iso}Substeps associated with the $\sqrt{3}\times \sqrt{3}$ commensurate phase formation in the nitrogen adsorption isotherm measurements at $T=77.4$~K on our ZYX substrate~(\opencircle) with a dummy cell and those on Grafoil substrate~(\fullcircle)~\cite{niimiRSI}. The intersections of the dashed lines show the ends of the substeps. The inset  additionally shows those at $T=71.8$~K~(\opensquare), $75.5$~K~$(\opentriangle)$ after mounting the ZYX substrate on our system with appropriate horizontal scaling.}
\end{minipage} 
\end{figure}

The N$_{2}$ monolayer capacity was measured at $T=71.8$~K,  $75.5$~K and $77.4$~K.
Substeps associated with the $\sqrt{3}\times \sqrt{3}$ commensurate phase formation are shown in Figure \ref{fig:iso}. It shows the substep on ZYX substrate is clearer and steeper than that on Grafoil and the inset shows that the shape of the substeps remained clear after packing the ZYX substrate into our cell.
The total surface area of our substrate was estimated as $30.7$~m$^{2}$ from the ends of the substeps and the specific surface area is consistent with previous measurements~\cite{niimiRSI}.
\section{Preliminary Results}
\begin{figure}[htbp]
\begin{center}
\includegraphics[width=0.80\textwidth]{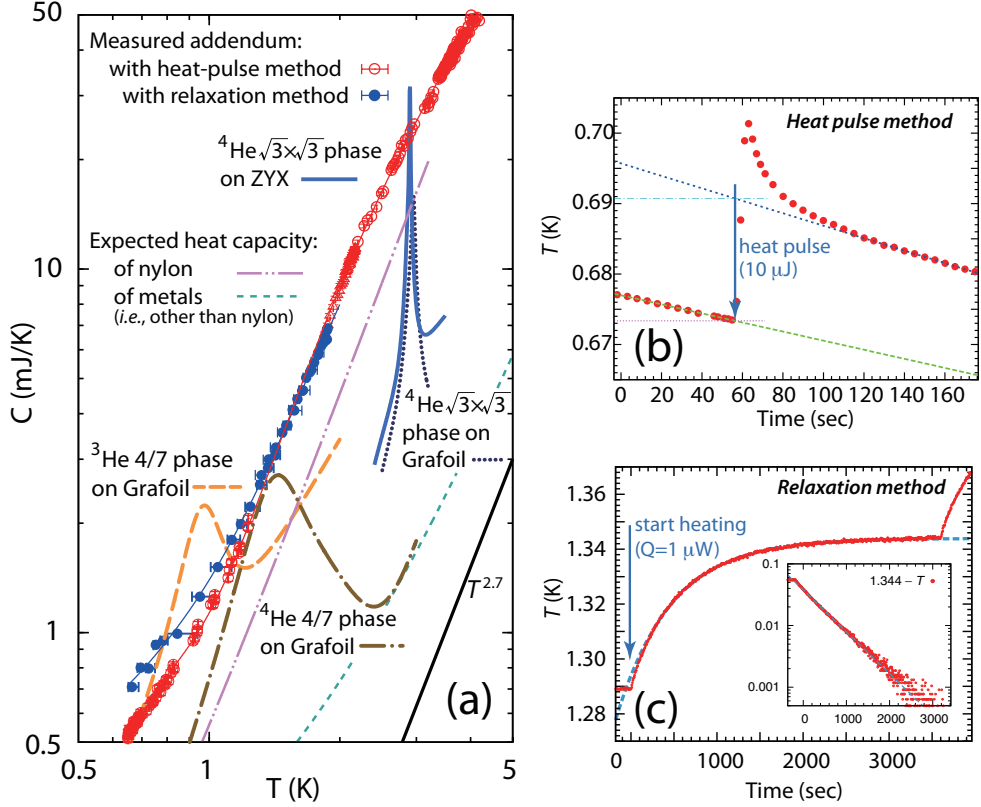}\hspace{0.03\textwidth}
\caption{\label{fig:c}{\bf(a)} A temperature dependency of the $C_{\mathrm{add}}$. Data measured with the adiabatic heat-pulse method (red circles), those with the relaxation method (blue disks), expected addendum of nylon(\dashddot) and that of other parts(\dashed) are shown. Expected heat-capacity of 2nd-layer  $^4$He(\chain)~\cite{grey}, $^3$He(\longbroken)~\cite{sv} in the 4/7 phases and 1st-layer $^4$He(\dotted) in the $\sqrt{3}\times \sqrt{3}$ phase on Grafoil and that on ZYX~\cite{root3} are also shown. {\bf(b)} A time evolution of the temperature of the addendum with the adiabatic heat-pulse method. {\bf(c)} That with the relaxation method.}
\end{center}
\end{figure}

Measured heat capacity of the empty calorimeter (addendum) is shown in Figure~\ref{fig:c}~(a).
The data points were taken by the adiabatic heat pulse method and 
the thermal relaxation method that is on a simple assumption that the sample is thermally connected only to the M/C plate whose temperature is fixed.
The measured addendum was about 2-times larger than expected in wide temperature range. It may because we used a monomer-cast nylon  (NIPPON POLYPENCO, MC901) instead of 6,6 nylon whose data we refer in the estimation.
The measured heat capacity of the empty cell was comparable with the anomalies of 2D Helium 4/7 phases on Grafoil. 2D helium on ZYX is generally expected to show bigger heat-capacity anomalies than that on Grafoil and it makes the measurements easier.

The typical measuring time with the heat-pulse method was several minutes (Figure~\ref{fig:c}~(b))
 and that with the relaxation method was 1~hour (Figure~\ref{fig:c}~(c)). 
The measured heat capacities of these two methods are systematically different below $1$~K because our cell is made of nylon with poor thermal conductivity below $1$~K. The measured heat capacity with the heat pulse method differed slightly by changing the time window from 2 to 10 minutes.

\section{Summary}
We have constructed a heat capacity measurement system to examine more widely spread 2D helium on ZYX graphite.
It will be possible to determine the critical exponents of the heat capacity anomaly of 2D helium 4/7 phases if the anomaly indicates an order-disorder transition and if the platelet size of ZYX substrate is wide enough.
Even if not, scaling behaviors are informative to determine the origin of the heat capacity anomalies and we expect to produce fruitful results soon.

\end{document}